\documentclass[11pt,english]{article}
\usepackage[T1]{fontenc}
\usepackage[utf8]{luainputenc}
\usepackage{amsmath}
\usepackage{amssymb}
\usepackage{setspace}
\usepackage{esint}
\makeatletter
\usepackage{babel}

\makeatother

\usepackage{babel}
\begin{document}

\title{Truncation of Einstein equations through Gravitational Foliation }
\author{{\small{}Merav Hadad{} }\thanks{{\small{}{}meravha@openu.ac.il }}}
%\author{Merav Hadad\footnote{meravha@openu.ac.il}\\	
%		\affiliation{Department of Natural Sciences, \\The Open University of Israel, Raanana 43107, Israel}\\
%	\affiliation{Astrophysics Research Center of the Open University (ARCO),\\ The Open University of Israel, P.O. Box 808,
%Ra’anana 4353701, Israel}}
\maketitle
\begin{abstract}
In previous works we suggested to consider a $(3+1)D$ quantum gravitational
field as an “evolution” of a $(2+1)D$ renormalized quantum gravitational
field along the direction of the gravitational force. The starting
point of the suggestion is derivation of a unique hypersurface which
looks effectively like $(2+1)D$ from the point of view of Einstein
equations in $(3+1)D$. In this paper we derive such unique hypersurfaces
for different kinds of static spherical metrics. We find that
these hypersurfaces exist whenever all the components of the gravitational
force field vanish on the hypersurface. We discuss the implication
of this result and the necessary further work.
\end{abstract}

\section{Introduction}

The conventional attempts to quantize the gravitational theory by means of Hamiltonian
and ADM formalism lead to a non renormalizable theory and to the
problem of time. Rather than giving up the powerful Hamiltonian formalism
when general relativity theories are concerned, we suggested \cite{Hadad:2019mga}
to use this formalism differently. We suggested to consider the symmetry
breaking caused by a gravitational force field, and to use the force
field direction as an independent parameter through which states evolve.
This mean that instead of singling out the direction of a time vector
field in the Hamiltonian formalism, we single out the direction of
the gravitational force.\footnote{We understand that the term “gravitational force field” is confusing since in gravity
	we rarely talk about the “gravitational force”. Here the term “gravitational force field”
	is used simply to refer to the acceleration field of a family of observers which do not change their spatial coordinates in a given coordinate system.} 

Though this suggestion is supported by several works \cite{merav,meravLevy,park2,park3,Davidson:2014tda,Ghaffarnejad},
which will be reviewed in the next section, its effect on causality
is unclear. To see this note that the direction of the foliation,
which is along the gravitational force field gives non-causal brackets
since the foliation is actually space-like directed and not time-like
directed. Recently we proved \cite{Hadad:2019xw} that under some
conditions, one can develop a causal quantum theory using space-like
directed foliation and it turns out that these conditions are useful
from the quantum gravity point of view. However, these conditions
are useful only if one can find, for a given metric background, a
kind of unique holographic hypersurface which looks effectively as
a $(2+1)D$ from the point of view of Einstein equations in $(3+1)D$.

The purpose of this paper is to find specific examples for such a
unique hypersurface in static spherically symmetric metrics. To
begin with we obtain the gravitational force field direction by considering
the acceleration vector field of static observers in different
kinds of spherically symmetric metrics. Next, by using the ADM formalism,
we foliate spacetime along this direction and find the conditions
needed for hypersurfaces to appear effectively as $(2+1)D$ from the
point of view of the Einstein equations. We found that these conditions
are fulfilled whenever all the components of the acceleration vector
field vanish on the hypersurface. 

Before discussing the implications of this result, let us first see
why this unique hypersurface is expected to be helpful for constructing
a causal $(3+1)D$ quantum gravity theory, even though its construction
involves \textquotedbl{}evolution\textquotedbl{} along a spatial direction.
As was shown in \cite{Hadad:2019xw}, Poisson brackets become non-causal
when foliating along a space like directed vector field. Thus the
only way to obtain the required causal classical field brackets for
this kind of foliation is by finding the fields' commutation relations
in some other way. In \cite{Hadad:2019xw} they were obtained from
the given theory on the hypersurface and taking them to their classical
limit. Moreover, since the hypersurface looks effectively a (2+1)D
from the point of view of the Einstein equations, a renormalized quantum
gravity theory can indeed be constructed on it. This construction
leads to a causal quantum gravitational theory in (3+1)D, even though
it \textquotedbl{}evolves\textquotedbl{} along a spatial direction. 

Now we examine the advantages of this construction and the significance
of our result. 

The advantage of constructing a gravitational theory using our unique
hyperspace is obvious. The unique hyperspace enables derivation of
quantum gravitational fields which are already renormalized on the
hypersurface. Whether the evolution of these specific fields \textquotedbl{}out
from the hypersurface\textquotedbl{} along the gravitational force
direction preserves their property of \textquotedbl{}being renormalizable\textquotedbl{}
remains to be seen. But the fact that the structure is based on a
renormalized quantum gravity looks promising and may lead the way
to construction of a (3+1)D renormalized quantum theory. 

Moreover, this contraction leads to a very interesting and important
result. It relates the acceleration of static observers in a given
coordinate system to the non-renormalizabilty of the quantum gravitational
theory. To see this connection note that our findings suggest that
when all the components of the acceleration vector field vanish, construction
of a renormalized gravitational theory is possible. In other words,
this relates the non-renormalizability property of the gravitational
theory to the existence of acceleration in a curved spacetime. This
relationship is predicted in \cite{merav}. We expand on this subject
in our conclusions.

Note that usually the Hamiltonian method is typically used to achieve a background-independent nonperturbative quantization. In our analysis, one uses the Hamiltonian method in order to limit the (3+1)D gravitational theory to an affective (2+1)D on a specific fixed background. Whereas the quantization of the affective (2+1)D gravitational theory can be done in various methods, we expect that in order to obtain a quantized (3+1)D gravitational theory, i.e. an extension to the forth direction, one should consider fluctuations around the background. Thus although we use the Hamiltonian method,  our analysis will not be a background-independent nonperturbative quantization.

The rest of the paper is organized as follows: Section 2 defines the
gravitational foliation force and discuss its implications. In section
2.1 we define and perform the gravitational foliation. In subsection
2.2 we discuss quantum gravity properties supporting foliation along
the gravitational force. In subsection 2.3 we deal with the expected
ambiguity regarding causality and its solution in the context of the
gravitational theory. In section 2.4 we derive the conditions for
hypersurfaces that appear effectively as $(2+1)D$ from the point
of view of the Einstein equations. In Section 3 we find these hypersurfaces
in different static spherical metrics. In section 3.1 we consider
the extremal black hole and show that although its horizon fulfills all the
necessary conditions, it can not be a proper candidate for the hypersurface to appear effectively as a $(2+1)D$ from the point of view of the Einstein equation. In section
3.2 we take a toy model and derive the condition in order for this
hypersurface to exist. In section 3.3 we find the condition for this
case for a general static spherical symmetric metric, and show
that it is fulfilled whenever all the components of the acceleration
vector field vanish on the hypersurface. In section 4 we discuss the
implications of our findings as well as possible future work. Section
5 is a summary. Finally, in the appendix, we derive the Einstein equation
when singling out a space like vector field direction instead of time
like vector field.

\section{Quantum gravity and gravitational foliation}

Obtaining a gravitational theory from microscopic objects or quantum
fields is an extremely important and challenging aim. Attempts to
describe gravity using microscopic objects, such as strings, loops
or triangles, have not yet led us to the desired Einstein equations.
Attempts to treat the gravitational metric as simply another quantum
field are problematic. Use of the Hamiltonian formalism in order to
quantize the gravitational theory leads to a non renormalizable theory
since it involves spin-2 massless fields and this kind of theory is
not renormalizable in more than $2+1$ dimensions. 

Instead of giving up the powerful Hamiltonian formalism where general
relativity theories are concerned, on the one hand, and in attempt
to use the renormalized gravitational theories on the $2+1$ dimensions,
on the other hand, we suggested \cite{Hadad:2019mga} a different
use of the Hamiltonian formalism. In our approach, we consider the
symmetry breaking caused by the gravitational force field for static
observers in a given coordinate system and use the direction of that
field as the direction through which states ``evolve.'' 

This approach is supported by several examples which relate the gravitational
foliation to different aspects of quantum gravity, on the one hand,
but is expected to be problematic since the direction of any force
field is space-like, and not time-like and thus leads to ambiguity
regarding causality. 

We begin by foliating along the gravitational force direction and
use this in order to rewrite the Einstein equation in the ADM formalism.
We then provide examples which relates this approach to different
aspects in quantum gravity. Next we deal with the causality issue
and its possible solution. We conclude this section by deriving the
necessary condition for the hypersurface to appear effectively as
$(2+1)D$ from the point of view of the Einstein equations.

\subsection{Gravitational foliation}

The first step is to define the gravitational force field direction.
For a given metric in a given coordinate system we calculate the gravitational
force direction for static observers. The 4-velocity vector field
of the static observer is $u^{a}=(\sqrt{-g^{00}},0,0,0)$ and
her 4-acceleration vector field is given by $a^{a}=u^{i}\nabla_{i}u^{a}$.
Thus the direction of the gravitational force field is given by $n^{a}=a^{a}/a$
where $a$ is the magnitude of the acceleration $a=\sqrt{a^{i}a_{i}}$.
Note that $n^{a}$ is a space like vector field.

Next we use the standard foliation of spacetime with respect to some
spacelike hypersurfaces whose directions are $n^{a}$. The lapse function
$N$ and shift vector $W_{a}$ satisfy $r_{a}=Nn_{a}+W_{a}$ where
$r^{a}\nabla_{a}r=1$ and $r$ is constant on $\Sigma_{\text{r}}$.
The $\Sigma_{r}$ hyper-surface metric $h_{ab}$ is given by $g_{ab}=h_{ab}+n_{a}n_{b}$.
The extrinsic curvature tensor of the hyper-surfaces is given by $K_{ab}=-\frac{1}{2}\mathcal{L}_{n}h_{ab}$
where $\mathcal{L}_{n}$ is the Lie derivative along $n^{a}$. Instead
of the$(3+1)D$ Einstein equations

\begin{equation}
R_{ab}^{(4)}=8\pi\left(T_{ab}-\frac{1}{2}Tg_{ab}\right)
\end{equation}
one finds (see appendix) a kind of $(2+1)D$ Einstein equations :
\begin{equation}
R_{ab}^{(3)}-KK_{ab}+2K_{ai}K_{b}^{i}+N^{-1}\left(\mathcal{L}_{r}K_{ab}-D_{a}D_{b}N\right)=8\pi\left(S_{ab}-\frac{1}{2}\left(S+P\right)h_{ab}\right),\label{eq:Estn3D}
\end{equation}
and the two constraints:
\begin{equation}
R^{(3)}-K^{2}+K_{ab}K^{ab}=-16\pi P,\label{eq:constrnt1}
\end{equation}
\begin{equation}
D_{a}K-D_{b}K_{a}^{b}=8\pi F_{a}.\label{eq:constrnt2}
\end{equation}
where $D_{a}$ represent the 2+1 covariant derivatives, $S_{ab}=h_{ac}h_{ad}T^{cd}$
, $P=n_{c}n_{d}T^{cd}$ and $\text{F}_{a}=-h_{ac}n_{b}T^{cb}$. We
will see that this foliation is relevant for several different areas
in physics which deal with the expected quantum gravity properties. 

\subsection{Background supporting foliation along the gravitational force}

We now provide some examples from our own work as well as that of
others, showing that gravitational foliation can be useful for aspects
of quantum gravity. The first example involves the surface density
of space time degrees of freedom (DoF). These are expected to be observed
by an accelerating observer in curved spacetime. This DoF surface
density was first derived by Padmanabhan \cite{Padmanabhan} for a
static spacetime using thermodynamic considerations. We found that
this density can also be constructed from specific canonical conjugate
pairs as long as they are derived in a unique way \cite{merav}. These
canonical conjugate pairs must be obtained by foliating spacetime
with respect to the direction of the gravitational vector force field.
Note that this aspect reinforces the importance of singling out a
very unique spatial direction: the direction of a gravitational force.

The second example involves string theory excitation. It was found
that some specific kind of singularity is obtained by string theory
excitations of a D1D5 black hole \cite{D1D5}. We found \cite{meravLevy}
that these singularities can also be explained using the uncertainty
principle, as long as the variables in the uncertainty principle are
obtained in a unique way: they must be canonical conjugate pairs which
are obtained by singling out the radial direction. The radial direction
can be regarded as the direction of a gravitational force for observers
that are static with respect to this coordinate system. Thus,
these singularities, which according to string theory are expected
in quantum gravity theories, are derived by the uncertainty principle
only when singling out the gravitational force direction. 

The third example involves “holographic quantization” which uses spatial
foliation in order to quantize the gravitational fields for different
backgrounds in Einstein theory. This is carried out by singling out
one of the spatial directions in a flat background,
and also singling out the radial direction for a Schwarzschild metric
\cite{park2}. Moreover, other works \cite{park3} even suggest that
the holographic quantization causes the (3+1)D Einstein gravity to
become effectively reduced to (2+1)D after solving the Lagrangian
analogues of the Hamiltonian and momentum constraints. 

The fourth example involves the developing of a quantum black hole
wave packet \cite{Davidson:2014tda}. In this case, the gravitational
foliation is used in order to obtain a quantum Schwarzschild black
hole, at the mini super spacetime level, by a wave packet composed
of plane wave eigenstates.

The fifth and final example involves the Wheeler-De Witt metric probability
wave equation. Recently, in \cite{Ghaffarnejad}, foliation in the
radial direction was used to obtain the Wheeler-De Witt equation on
the apparent horizon hypersurface of the Schwarzschild de Sitter black
hole. By solving this equation, the authors found that a quantized
Schwarzschild de Sitter black hole has a nonzero value for the mass
in its ground state. This property of quantum black holes leads to
stable black hole remnants.

Whereas our current approach relies on these examples, which relate
gravitational foliation to different aspects of quantum gravity, it
also strongly relies on the fact that it is possible to quantize a
$(2+1)D$ gravitational theory. Though a $(2+1)D$ gravitational theory
is believed to be a toy model for quantum gravity, we suggest that
a $(3+1)D$ theory may be regarded as a continuation along the gravitational
force field direction of a quantized $(2+1)D$ gravitational theory.
Given the fact that a renormalized $(2+1)D$ quantum gravity theory
can be obtained, this construction leads to a $(3+1)D$ quantum gravity
originated from a renormalized theory. Whether or not this construction
leads to an effectively renormalized gravitational theory on the $(3+1)D$
remains to be seen. However, we argue that even if this way of construction
does not lead to a renormalized quantum gravitational field theory
in the $(3+1)D$ but leads to the correct Einstein field equations,
this construction can be considered as a proof that our inability
to renormalized the $(3+1)D$ theory is related directly to the acceleration
relative to a given coordinate system. We expand this subject in section
4. 

Our suggestion leads to a $(3+1)D$ gravitational theory which “evolves”
along the gravitational force direction, i.e. an evolution along a
space like directed vector field. This leads to the causality vagueness.

\subsection{The causality ambiguity and its solution}

In our approach, it is necessary to single out the direction of the
gravitational force vector field instead of the direction of time
like vector field. Since the direction of any force field is space-like,
and not time-like this suggestion leads to a lack of clarity regarding
the basic concepts of relativistic quantum field theories: causality,
probability, conservation, unitarity and more. In order to overcome
these anticipated difficulties, we investigated the outcome of such
foliation on relativistic free scalar fields. We found that, under
some conditions, one can derive a causally quantum theory using non-Cauchy
foliation. However, it seems that the main problem is contraction
of the causal fields brackets on the non Cauchy hypersurface, since
the usual Poisson brackets are not useful when using non-Cauchy foliation. 

At this point it is not clear how to obtain such \textquotedbl{}Poisson-like\textquotedbl{}
brackets using the conventional mathematical definition, and we need
to derive them in some other way. Therefore we propose to derive the
quantum commutation relations between the fields on the unique non-Cauchy
hypersurface and then use these to obtain the classical brackets. 

Though in general this restriction seems problematic, from the quantum
gravity point of view it turns out to be promising. The main reason
is that in some cases \cite{carlip} we do know how to quantize a
$(2+1)D$ gravitational theory.\footnote{Note that a  (2 + 1)-dimensional gravity is
	tricky to quantize: for example a partition function for 3D pure
	gravity with a negative cosmological constant was derived in \cite{Maloney:2007ud}
	and \cite{Keller:2014xba} and exhibit several problematic features, including a
	negative density of states in certain regimes.} Thus for a given metric background
in a $(3+1)D$ gravitational theory, we can look for unique hypersurfaces
that appear effectively as $(2+1)D$ from the point of view of the
Einstein equations. If we manage to do so, we can construct a renormalized
quantum gravity on this unique $(2+1)D$ hypersurface and obtain the
commutation relations of the quantum gravitational fields on the hypersurface.
In this way, we can easily deduce the causal classical brackets of
the fields without using the Poisson brackets. Thus, as was found
in the scalar case \cite{Hadad:2019xw}, one can derive the Hamilton-like
equations along the hypersurface direction and use these classical
causal brackets in order to obtain the causal quantum gravitational
theory in $(3+1)D$. We expand this subject in section 4.

This construction of quantum gravity in $(3+1)D$ relies on known
$(2+1)D$ renormalized gravitational theory on a unique hypersurface.
In the next subsection, we define the unique hypersurface that truncates
the Einstein equations and find the condition for it to generate an
effectively $(2+1)D$ Einstein equations.

\subsection{Truncation of Einstein equations: the condition for effectively $(2+1)D$
Einstein equations on the hypersurface}

In the first subsection we found that the gravitational foliation
gives a kind of $(2+1)D$ Einstein equations: 
\[
R_{ab}^{(3)}-KK_{ab}+2K_{ai}K_{b}^{i}+N^{-1}\left(\mathcal{L}_{r}K_{ab}-D_{a}D_{b}N\right)=8\pi\left(S_{ab}-\frac{1}{2}\left(S+P\right)h_{ab}\right),
\]
and the two constraints:
\[
R^{(3)}-K^{2}+K_{ab}K^{ab}=-16\pi P,
\]
\[
D_{a}K-D_{b}K_{a}^{b}=8\pi F_{a}.
\]
Thus if we able to find a unique hypersurface $r=r_{0}$ so that:

\begin{equation}
B_{ab}\equiv KK_{ab}-2K_{ai}K_{b}^{i}-N^{-1}\left(\mathcal{L}_{r}K_{ab}-D_{a}D_{b}N\right)=0\label{eq:IsThisZiro}
\end{equation}
then we have: 
\begin{equation}
R_{ab}^{(3)}=8\pi\left(S_{ab}-\frac{1}{2}\left(S+P\right)h_{ab}\right),
\end{equation}
which is almost the Einstein equations in 2+1 dimension, when $P$
serves as a cosmological constant. However, note that the Einstein
equations in $(2+1)D$ are $R_{ab}^{(3)}=8\pi\left(S_{ab}-Sh_{ab}\right)$
and thus we have only obtained an ``Einstein-like equation.'' Only
if $S=P$ do we get exactly the expected Einstein equation in $(2+1)D$
. Thus, for example, we do not expect the conservation of the energy
momentum $S_{ab}$ to hold on this hypersurface. Moreover, one has
to consider the fact that in (2+1)D and (3+1)D gravitational constant
and even the energy-momentum tensor have different units.

This unique hypersurface is interesting. Although we do not know how
to obtain a renormalized quantum gravity theory in 3+1 dimensions,
a renormalized quantum theory in 2+1 dimensions can nevertheless be
obtained \cite{carlip}. Thus, when the $(3+1)D$ Einstein equations
reduce to $(2+1)D$ Einstein-like equations on some hypersurface $r=r_{0}$,
we can quantize the gravitational fields on the hypersurface $r=r_{0}$
at least with respect to this foliation. In the next section we find
this kind of unique hypersurface for different static spherical
metrics. 

\section{Spherical symmetry examples}

In this section we deal with different examples of truncation hypersurfaces.
All of them assume static spherically symmetric metrics\footnote{In general, given an arbitrary vector field on a spacetime, there
will not exist hypersurfaces normal to the vector field. For a family of surfaces normal to $a^a$ to exist,
Frobenius’ theorem requires that the vector field $a^a$ must be hypersurface orthogonal. In special cases with a
high degree of symmetry (like the static, spherically symmetric cases), the symmetries can ensure that $a^a$ is indeed hypersurface orthogonal
and therefore the foliation exists}. Note that although one may expect that it is useful to first tackle a pure gravity system, we will see that our analysis becomes easier when dealing with a gravity-matter system.

\subsection{First example: extremal black hole}

We start with the metric of an extremal black
hole and show that although its horizon fulfills all the
necessary conditions, this metric can not be a proper candidate for the hypersurface to appear effectively as a $(2+1)D$ from the point of view of the Einstein equation. 

In this case 
\[
ds^{2}=-\left(1-\frac{M}{r}\right)^{2}dt^{2}+\left(1-\frac{M}{r}\right)^{-2}dr^{2}+r^{2}d\varOmega^{2}.
\]
The 4-velocity and 4-acceleration vector fields for a static observer
in this metric are 
\[
u^{a}=(\left(1-\frac{M}{r}\right)^{-1},0,0,0)
\]
\[
a^{a}=(0,\frac{M}{r^{2}}(1-\frac{M}{r}),0,0),
\]
The direction of acceleration is 
\[
n^{a}=(0,(1-\frac{M}{r}),0,0).
\]
The induced metric becomes
\[
h_{ab}=\left(\begin{array}{cccc}
\left(1-\frac{M}{r}\right)^{2} & 0 & 0 & 0\\
0 & 0 & 0 & 0\\
0 & 0 & r^{2}\\
0 & 0 & 0 & r^{2}sin^{2}\theta
\end{array}\right)
\]
and the extrinsic curvature

\[
K_{ab}=\left(\begin{array}{cccc}
-\frac{M\left(r-M\right)^{2}}{r^{4}} & \times & 0 & 0\\
\times & \times & \times & \times\\
0 & \times & \left(r-M\right)\\
0 & \times & 0 & \left(r-M\right)sin^{2}\theta
\end{array}\right).
\]
The Lie derivative along $r^{a}$ of the extrinsic curvature is 

\[
\mathcal{L}_{\text{r}}K_{ab}=\left(\begin{array}{cccc}
2\frac{M^{2}}{r^{5}}\left(1-\frac{M}{r}\right)^{2}\left(1-\frac{2M}{r}\right) & \times & 0 & 0\\
\times & \times & \times & \times\\
0 & \times & \frac{M}{r^{2}}\left(1-\frac{M}{r}\right)\\
0 & \times & 0 & \frac{M}{r^{2}}\left(1-\frac{M}{r}\right)sin^{2}\theta
\end{array}\right)
\]
and since $N=\left(1-\frac{M}{r}\right)^{-1}$ and $W_{a}=0$ we obtain
\[
\begin{array}{ccc}
D_{0}D_{0}N & = & \frac{M^{2}}{r^{4}}\left(1-\frac{M}{r}\right)\\
D_{2}D_{2}N & = & \frac{M}{r}\\
D_{3}D_{3}N & = & \frac{M}{r}sin^{2}\theta\text{.}
\end{array}
\]
 Finally, we have everything we need in order to calculate the truncation
tensor $B_{ab}$ defined in eq. (\ref{eq:IsThisZiro}). It turns out
that all its components vanish on the horizon: $r=M$, and thus we
conclude that the term in Einstein equations that depends on the extrinsic
curvature vanishes on the horizon: $r=M$ and eq. (\ref{eq:IsThisZiro})
holds. This means that from the point of view of Einstein equations, the
hypersurface denoted by $r=M$ does not \textquotedbl{}feel\textquotedbl{}
the radial direction and we may expect that effectively the Einstein equations can be
described as leaving on $(2+1)D$. However, note that in this case the time-time component of the induced metric becomes $h_{00}=0$ and thus this surface is not a good candidate for Einstein equations leaving on $(2+1)D$.

\subsection{Second example: toy model}

It turns out that although the extremal black hole is not a good candidate for our suggestion, one can find
examples of gravitational foliation that lead to an effectively
$\left(2+1\right)D$ theory from the point of view of the Einstein
equations. For example we consider the metric

\[
ds^{2}=-\left(1-\frac{A^{2}}{r^{2}}+\frac{B^{3}}{r^{3}}\right)dt^{2}+\left(1-\frac{A^{2}}{r^{2}}+\frac{B^{3}}{r^{3}}\right)^{-1}dr^{2}+r^{2}d\varOmega^{2}
\]
In this case the 4-velocity and 4-acceleration vector fields for a
static observer in this metric are 
\[
u^{a}=\left(\left(1-\frac{A^{2}}{r^{2}}+\frac{B^{3}}{r^{3}}\right)^{-1/2},0,0,0\right).
\]
\[
a^{a}=\left(0,\frac{2Ar-3B^{2}}{2r^{4}},0,0\right),
\]
Note that all the components to the accelerating vector field vanish
on $r=3B^{2}/2A$.

Foliating spacetime along the direction of the acceleration vector
field, gives the induced metric 
\[
h_{ab}=\left(\begin{array}{cccc}
\left(1-\frac{A^{2}}{r^{2}}+\frac{B^{3}}{r^{3}}\right) & 0 & 0 & 0\\
0 & 0 & 0 & 0\\
0 & 0 & r^{2} & 0\\
0 & 0 & 0 & r^{2}sin^{2}\theta
\end{array}\right)
\]
and the extrinsic curvature of hypersurfaces directed along the acceleration
vector field is

\[
K_{ab}=\sqrt{(3B^{2}-2Ar)^{2}(B^{2}-Ar+r^{3})}\left(\begin{array}{cccc}
-\frac{1}{2}r^{-11/2} & \times & 0 & 0\\
\times & \times & \times & \times\\
0 & \times & r^{-3} & 0\\
0 & \times & 0 & r^{-3}sin^{2}\theta
\end{array}\right).
\]
The Lie derivative along $r^{a}$ of the extrinsic curvature vanishes
everywhere except the following components: 
\[
\begin{array}{ccc}
\mathcal{L}_{r}K_{00} & = & \sqrt{\frac{(3B^{2}-2Ar)^{2}}{r^{5}(B^{2}-Ar+r^{3})}}\frac{-33B^{4}-24B^{2}r(-2A+r^{2})+4Ar^{2}(-4A+3r^{2})}{8r^{8}}\\
\mathcal{L}_{r}K_{22} & = & \sqrt{\frac{(3B^{2}-2Ar)^{2}}{r^{5}(B^{2}-Ar+r^{3})}}\frac{(B^{2}-2r^{3})}{4r^{3}}\\
\mathcal{L}_{r}K_{33} & = & \sqrt{\frac{(3B^{2}-2Ar)^{2}}{r^{5}(B^{2}-Ar+r^{3})}}\frac{(B^{2}-2r^{3})}{4r^{3}}sin^{2}\theta
\end{array}
\]
and since $N=\left(1-\frac{A^{2}}{r^{2}}+\frac{B^{3}}{r^{3}}\right)^{-1/2}$
we obtain 
\[
\begin{array}{ccc}
D_{0}D_{0}N & = & \frac{\left(3B^{2}-2Ar\right)^{2}}{2r^{7}\sqrt{B^{2}-Ar+r^{3}}}\\
D_{2}D_{2}N & = & \frac{3B^{2}-2Ar}{2r^{2}\sqrt{B^{2}-Ar+r^{3}}}\\
D_{3}D_{3}N & = & \frac{3B^{2}-2Ar}{2r^{2}\sqrt{B^{2}-Ar+r^{3}}}sin^{2}\theta
\end{array}
\]
Finally, we have everything we need in order to calculate the truncation
tensor $B_{ab}$ defined in eq (\ref{eq:IsThisZiro}). As one might
expect, eq. (\ref{eq:IsThisZiro}) holds when the component of the
acceleration vector field vanishes, as in the extremal black hole
case. Since $a^{a}=\left(0,\frac{2Ar-3B^{2}}{2r^{4}},0,0\right)$
we find that the acceleration vector field vanishes on $r=\frac{3B^{2}}{2A}$.
Note that since for general $A$ and $B$ the term $f(r)_{r=\frac{3B^{2}}{2A}}=\left(1-\frac{4A^{3}}{27B^{4}}\right)$
does not vanish on $r=\frac{3B^{2}}{2A}$, the hypersurface denoted
by $r=\frac{3B^{2}}{2A}$ is not a horizon.

It is interesting to note that in this example the energy momentum
tensor $S_{b}^{a}$ and momentum $P$ on the hyper-surface $r=\frac{3B^{2}}{2A}$
are 
\begin{equation}
S_{b}^{a}=\left(\begin{array}{ccc}
-\frac{16A^{5}}{243B^{8}} & 0 & 0\\
0 & \frac{16A^{5}}{81B^{8}} & 0\\
0 & 0 & \frac{16A^{5}}{81B^{8}}
\end{array}\right),\textrm{ }P=-\frac{16A^{5}}{243B^{8}}.\label{eq:general P}
\end{equation}
 and thus it does not represent an AdS universe. 

\subsection{Third example: general static spherical metric}

We can now investigate the conditions that are required in order that
a hypersurface on $\left(3+1\right)D$ spherical static universe
can be regarded as a $\left(2+1\right)D$ gravitational theory. In
order to do so, we consider a general static spherically symmetric
metric 
\[
ds^{2}=-f(r)dt^{2}+f^{-1}(r)dr^{2}+r^{2}d\varOmega^{2}.
\]
In this case the 4-velocity and 4-acceleration vector fields for a
static observer are 
\[
u^{a}=\left(f^{-1/2}(r),0,0,0\right),
\]
\[
a^{a}=\left(0,\frac{1}{2}f',0,0\right).
\]
and so $N=f^{-1/2}(r)$ , $W_{a}=0$ and the direction of the acceleration
is always radial. Thus foliating spacetime along the direction of
the acceleration vector field gives the induced metric 
\[
h_{ab}=\left(\begin{array}{cccc}
f(r) & 0 & 0 & 0\\
0 & 0 & 0 & 0\\
0 & 0 & r^{2} & 0\\
0 & 0 & 0 & r^{2}sin^{2}\theta
\end{array}\right)
\]
and the extrinsic curvature of hypersurfaces directed along the acceleration
vector field is

\[
K_{ab}=\sqrt{f(r)}\left(\begin{array}{cccc}
-\frac{1}{2}f' & \times & 0 & 0\\
\times & \times & \times & \times\\
0 & \times & r & 0\\
0 & \times & 0 & rsin^{2}\theta
\end{array}\right).
\]
The Lie derivative along $r^{a}$ of the extrinsic curvature vanishes
everywhere except the following components: 
\[
\begin{array}{ccc}
\mathcal{L}_{r}K_{00} & = & -\frac{1}{8}\frac{f'}{\sqrt{f}}\left(f'^{2}+2ff''\right)\\
\mathcal{L}_{r}K_{22} & = & \frac{1}{4}\frac{f'}{\sqrt{f}}\left(rf'+2f\right)\\
\mathcal{L}_{r}K_{33} & = & \frac{1}{4}\frac{f'}{\sqrt{f}}\left(rf'+2f\right)sin^{2}\theta
\end{array}
\]
and since $N=f^{-1/2}(r)$ we get 
\[
\begin{array}{ccc}
D_{0}D_{0}N & = & \frac{1}{4}\frac{f'^{2}}{\sqrt{f}}\\
D_{2}D_{2}N & = & -\frac{1}{2}\frac{rf'}{\sqrt{f}}\\
D_{3}D_{3}N & = & -\frac{1}{2}\frac{rf'}{\sqrt{f}}sin^{2}\theta
\end{array}
\]
 Thus the truncation tensor gives 
\[
B_{ab}=f'\left(\begin{array}{cccc}
\frac{1}{8}f'^{2}+\frac{1}{4}f''f-\frac{1}{r}f & \times & 0 & 0\\
\times & \times & \times & \times\\
0 & \times & r-\frac{1}{2}f-\frac{1}{4}rf' & 0\\
0 & \times & 0 & \left(r-\frac{1}{2}f-\frac{1}{4}rf'\right)sin^{2}\theta
\end{array}\right).
\]
As we see, the conditions of equation (\ref{eq:IsThisZiro}) hold
on a hypersurface $r=r_{0}$ if $f'\left(r_{0}\right)=0$ or if $\left(\frac{1}{8}f'^{2}+\frac{1}{4}f''f-\frac{1}{r}f\right)_{r=r_{0}}=\left(r-\frac{1}{2}f-\frac{1}{4}rf'\right)_{r=r_{0}}=0$.
Note that when the condition $f'\left(r_{0}\right)=0$ holds, then
the acceleration vector field $a^{a}=\left(0,\frac{1}{2}f',0,0\right)$
vanishes on $r=r_{0}$ and we see that the hypersurface does not have
to be a horizon.

The term $K^{2}-K_{ab}K^{ab}$ from the first constraint (\ref{eq:constrnt1})
equals to
\[
K^{2}-K_{ab}K^{ab}=\frac{2}{r^{2}}\left(f+rf'\right).
\]
As expected from the second example, this term does not vanish when
equation (\ref{eq:IsThisZiro}) holds. 

It is interesting to note that just as in eq. (\ref{eq:general P}),
in this example the energy momentum tensor $S_{b}^{a}$ and momentum
$P$ on the hyper-surface $r=r_{0}$ are 
\[
S_{b}^{a}=\left(\begin{array}{ccc}
\frac{1}{r^{2}}\left(f+rf'-1\right) & 0 & 0\\
0 & \frac{1}{r}f'+\frac{1}{2}f'' & 0\\
0 & 0 & \frac{1}{r}f'+\frac{1}{2}f''
\end{array}\right),\textrm{ }P=\frac{1}{r^{2}}\left(f+rf'-1\right).
\]
 and thus it does not represent an AdS universe. Moreover, note that at least for  a static spherically symmetric metrics our analysis becomes easier whenever one  deals with a gravity-matter system. Since for a static spherically symmetric metrics one can get a pure gravity system only when $\frac{1}{r^{2}}\left(f+rf'-1\right)=\frac{1}{r}f'+\frac{1}{2}f''=0$. \footnote{We do not believe that the consistent with the
 	usual energy conditions in general relativity, like the null energy condition, the weak
 	energy condition, are relevant in this surface.}

\section{Further work}

In this paper, we proved that under some conditions there exists a
unique hypersurface that causes the Einstein equations to look like
$(2+1)D$ . But in order to construct a renormalized quantum gravity
this is not sufficient. 

To start with, note that in this approach the "evolution" is not along "time" but along the spatial coordinate. This means that all the dynamics must be encoded on our unique hypersurface in advance.  In other words, the hypersurface must be holographic.
The holographic principle is not new \cite{tHooft:1993dmi,Susskind:1994vu}. According to this principle, the number of degrees of freedom fundamentally
scales like the area of surfaces and not the enclosed volume
as one would expect from local field theory. In order to extend this idea to any spacetime, the Bousso bound \cite{Bousso:1999cb} was formulated.  Next by defining two kinds of holographic screens (future and past), Bousso and Engelhardt \cite{Bousso:2015mqa,Bousso:2015qqa} proved an extended new area law. Recently, in \cite{Ben-Dayan:2020pbg} we related the entropy of any holographic
screen to the phase space entropy derived in \cite{merav} and by using the new area law we identified uniquely
the foliation direction needed for any given holographic screen.  Using this explicit foliation, the next step should be deriving the truncation of Einstein equations for the holographic screens defined by Bousso and Engelhardt.

Next, it is probably necessary to derive a Hamilton-like ADM equation
for the evolution along the direction of the gravitational force.
Except the constraint, this is expected to look like

\begin{equation}
\mathcal{L}_{r}\varPi_{ab}=-\left\{ \tilde{H},h_{ab}\right\} ,\textrm{ }\mathcal{L}_{r}h_{ab}=\left\{ \tilde{H},\varPi_{ab}\right\} \label{hamiltton fq1}
\end{equation}
 where 
\begin{align*}
L & =N\sqrt{-h}\left(^{(3)}R+K_{ij}K^{ij}-K^{2}\right)\\
\varPi^{ab}: & =\frac{\partial L}{\partial\mathcal{L}_{r}h_{ab}}=\sqrt{-h}\left(Kh^{ab}-K^{ab}\right)\\
\tilde{H}= & \varPi^{ab}\mathcal{L}_{r}h_{ab}-L\\
= & \sqrt{-h}\left[N\left(-^{(3)}R+h^{-1}\Pi_{ij}\varPi^{ij}-\frac{1}{2}h^{-1}\Pi^{2}\right)-2W_{i}D_{j}\left(h^{-1/2}\Pi^{ij}\right)\right]
\end{align*}
\\
Note that the brackets (\ref{hamiltton fq1}) are not trivial, and
must be identified correctly. This happens because when using spatially
directed foliation, instead of the usual time like foliation, one
can no longer use the Poisson brackets. Thus it is necessary to derive
a kind of \textquotedbl{}Poisson-like'' brackets between the classical
fields. In that case we would be able to obtain the relations

\[
\left\{ h_{ij},h_{ab}\right\} _{r=r_{0}},\left\{ h_{ij},\Pi_{ab}\right\} _{r=r_{0}},\left\{ \Pi_{ij},\Pi_{ab}\right\} _{r=r_{0}}.
\]

Our construction is helpful in order to derive the first one:$\left\{ h_{ij},h_{ab}\right\} _{r=r_{0}}$.
These brackets are known whenever one constructs a quantum gravity
theory on the unique $\left(2+1\right)D$ hypersurface. Thus the next
step upon quantizing $\left(2+1\right)D$ should be a derivation of
the quantum gravitational theory on the unique hypersurfaces specified
above. In order to do that note first that in $(2+1)D$ and $(3+1)D$
both the gravitational constant and the energy-momentum tensor have
different units.

In order to derive the second $\left\{ h_{ij},\varPi_{ab}\right\} _{r=r_{0}}$
and the third $\left\{ \varPi_{ij},\varPi_{ab}\right\} _{r=r_{0}}$a
different approach should be taken. For example, one may consider
the use of Peierls bracket \cite{Peierls} which is a more covariant
structure equivalent to the Poisson bracket but which can be built
directly from advanced and retarded Green’s functions for the linearized
equations of motion. 

Moreover, the proof that one can derive the Hamilton-like
equations by foliating spacetime along spatial direction were proven only for scalars \cite{Hadad:2019xw}. It seems that this result can easily be extended for any vector fields if one ignores complication associated with gauge invariance and work directly
with physical components. In this case, the action of each physical component will be the same as for a scalar field. For example, though in (3 + 1)D the metric has 10 components 8 of them are non-physical, and each of the two remaining physical components has an effective action of a scalar field. Thus, quantizing a vector field by foliating spacetime along the spatial direction is also possible whenever the vector field components have causal commutation relation on the non-Cauchy hypersurface. However, this result is relevant only when one ignores complications associated with gauge invariance in the Minkowski frame. Thus, the next step should be considering the implications of the gauge invariance on the derivation of causal quantum theory when singling out a spatial direction. Moreover, in order to extend our intuitiveness for gravity, one should also examine the derivation of causal quantum theory in the Rindler metric (for scalars and vector fields) when singling out a unique spatial direction: the acceleration direction.

\section{Summary and discussion}

In this paper, we used gravitational foliation in order to find a
few static spherical symmetric examples of hypersurfaces that
truncate the Einstein equations. We began with a derivation of the
Einstein equation in the ADM formalism when foliating along the gravitational
force direction. In order to find this direction we considered the
direction of the acceleration vector field of static observers
relative to the given coordinate system. Then we derived the conditions
for hypersurfaces that appear effectively $(2+1)D$ from the point
of view of the Einstein equations. Finally, we found these hypersurfaces
in different static spherical metrics. We found that the conditions
in this case are fulfilled whenever all the components of the acceleration
vector field vanish on the hypersurface.

Now further work is necessary. The next step should be to construct
a renormalized quantum gravity on this unique $(2+1)D$ hypersurface
and to obtain the commutation relations of the quantum gravitational
fields on the hypersurface. In this way, we could easily deduce the
causal classical brackets of the fields on this unique $(2+1)D$ hypersurface,
without using the classical Poisson brackets.

The advantage of constructing a gravitational theory using the unique
hyperspace is obvious. The unique hyperspace enables derivation of
quantum gravitational fields which are already renormalized on the
hypersurface. i. e. to obtain $\left\{ h_{ij},h_{ab}\right\} _{r=r_{0}}$
.Whether the evolution of these specific fields \textquotedbl{}out
from the hypersurface\textquotedbl{} along the gravitational force
direction keeps their feature of \textquotedbl{}being renormalizable\textquotedbl{}
remains to be seen. But the fact that the structure is based on a
renormalized quantum gravity looks promising and may lead the way
to construct a $(3+1)D$ renormalized quantum theory. 

Note that this procedure may be related to a kind of holography. In
our suggested formalism, the evolution of the gravitational fields
along the acceleration direction is determined by $\left\{ h_{ij},h_{ab}\right\} _{r=r_{0}},\left\{ h_{ij},\Pi_{ab}\right\} _{r=r_{0}},\left\{ \Pi_{ij},\Pi_{ab}\right\} _{r=r_{0}}$
which plays the role of the \textquotedbl{}initial\textquotedbl{}
or surface condition on the non-Cauchy hyper-surface $r=r_{0}$. This
construction gives all the information which encoded on the hyper-surface
and is needed to describe the evolution of the gravitational field
along the acceleration direction. However, whether this construction
gives all the information in the balk is remain to be seen. 

Moreover, this construction leads to a very interesting result since
it relates the non-renormalizabilty of the quantum gravitational theory
to acceleration in curved spacetime. To see this, note that our conditions
on the hypersurface are fulfilled whenever all the components of the
acceleration vector field of static observers vanish. Moreover,
these conditions cause our hypersurface to look effectively $(2+1)D$
from the point of view of the Einstein equation. This means that if
we find, for a given metric, a coordinate system that leads to the
vanishing of the acceleration components for all the hypersurfaces
directed along the acceleration vector field of static observers,
construction of a renormalized $(3+1)D$ gravitational theory could
be obtained. This happens because in this case, all hypersurfaces
look effectively like $(2+1)D$ from the point of view of the Einstein
equations. Since accelerated observers must use the $(3+1)D$ Einstein
equations, this re-normalized quantization cannot be applied for accelerated
observers. This may suggest that acceleration in curved spacetime
and the non-renormalizabilty of the quantum gravitational theory are
connected.

Not surprisingly, foliation along a vanishing acceleration vector
field is impossible. To see this note that if indeed all the components
of the acceleration vector field vanish everywhere, we deal with freely
falling observers. In this case, the direction of the acceleration
cannot be defined and our suggested foliation cannot be done. However,
this makes us wonder whether we may expect that freely falling observers
in a $(3+1)D$ can only use a $(2+1)D$ coordinate system in order
to describe the universe. This case is very interesting because this
suggests that at least freely falling observers may be able to obtain
a renormalized quantum gravitational theory by choosing the right
coordinate system. This idea is supported by \cite{merav,Padmanabhan}
which relates the extra gravitational degrees of freedom seen by static
observers to their acceleration in generalized theories of gravity.
Moreover, it was found that these extra gravitational degrees of freedom
vanish whenever the observers move on a geodesic. This reinforces
our motivation to investigate the possibility that freely falling
observers “see” less DoF and thus can renormalize the gravitational
theory. This suggestion needs further investigation. First: Are our
findings relating the vanishing of the acceleration components on
the hypersurface to effectively $(2+1)D$ Einstein equations relevant
even for non-spherically symmetric systems? What is the physical meaning
of a system of coordinates that is relevant only for freely falling
observers? Even if a $(2+1)D$ system of coordinates that is relevant
only for freely falling observers exists, how do we extend it for
accelerated observers which must use $(3+1)D$?

\textbf{Acknowledgments:} This research was supported by The Open
University of Israel's Research Fund (grant no. 510503).

\section*{Appendix: The ``spatial ADM'' formalism and derivation of Einstein
equations for spacial foliation}

When foliating along a space like vector field, instead of a time
like vector field, the Einstein field equations in the ADM formalism
slightly change. In this appendix we recalculate the Einstein equations
in the ADM formalism using a space like vector field, instead of time
like vector field.

We begin by considering the standard foliation of spacetime with respect
to some spacelike hypersurfaces whose directions are $n^{a}$. The
lapse function $N$ and shift vector $W_{a}$ satisfy $r_{a}=Nn_{a}+W_{a}$
where $r^{a}\nabla_{a}r=1$ and $r$ is constant on $\Sigma_{\text{r}}$(Thus
$n_{a}=N\nabla_{a}r$). The $\Sigma_{r}$ hyper-surfaces metric $h_{ab}$
is given by $g_{ab}=h_{ab}+n_{a}n_{b}$. The extrinsic curvature tensor
of the hyper-surfaces is given by $K_{ab}=-\frac{1}{2}\mathcal{L}_{n}h_{ab}$
where $\mathcal{L}_{n}$ is the Lie derivative along $n^{a}$. \footnote{The intrinsic curvature $R_{ab}^{(3)}$ is then given by the 2+1 Christoffel
symbols: $\Gamma_{ab}^{k}=\frac{1}{2}h^{kl}\left(\frac{\partial h_{lb}}{\partial x^{a}}+\frac{\partial h_{al}}{\partial h^{b}}-\frac{\partial h_{ab}}{\partial x^{l}}\right)$
so that $R_{ab}^{(3)}=\frac{\partial\Gamma_{ab}^{k}}{\partial x^{k}}-\frac{\partial\Gamma_{ak}^{k}}{\partial x^{b}}+\Gamma_{ab}^{k}\Gamma_{kl}^{l}-\Gamma_{al}^{l}\Gamma_{lb}^{k}$
.} 

We use this foliation and rewrite the$(3+1)D$ Einstein equations 

\begin{equation}
^{(4)}R_{ab}=8\pi\left(T_{ab}-\frac{1}{2}Tg_{ab}\right).\label{eq:4d einst eq}
\end{equation}
in terms of the induced metric. The starting point of the calculation
(see for example see \cite{book 3+1}) is the Gauss relation which
we derive here for non-Cauchy foliation from the Ricci identity on
the $\Sigma_{r}$ hyper-surfaces:
\begin{equation}
D_{a}D_{b}v^{c}-D_{b}D_{a}v^{c}={}^{(3)}R_{mab}^{c}v^{m}\label{eq: ricci}
\end{equation}
where $v^{m}$ is a generic vector field tangent to $\Sigma_{r}$.
Relating the $D$-derivative to the $\nabla$-derivative and using
$\nabla_{m}h_{ab}=\nabla_{m}(g_{ab}-n_{a}n_{b})=-\nabla_{m}n_{a}n_{b}-n_{a}\nabla_{m}n_{b}$,
$h_{b}^{n}n_{n}=0$ and $h_{a}^{m}h_{b}^{n}\nabla_{m}n_{n}=-K_{ab}$
one gets: 
\[
\begin{array}{ccc}
D_{a}D_{b}v^{c} & = & h_{a}^{m}h_{b}^{n}h_{i}^{c}\nabla_{m}\left(h_{n}^{s}h_{l}^{i}\nabla_{s}v^{l}\right)\\
 & = & h_{a}^{m}h_{b}^{n}h_{i}^{c}(-n^{s}\nabla_{m}n_{n}h_{l}^{i}\nabla_{s}v^{l}-h_{n}^{s}\nabla_{m}n^{i}\underset{=-v^{i}\nabla_{s}n_{l}}{\underbrace{n_{l}\nabla_{s}v^{l}}}+h_{n}^{s}h_{l}^{i}\nabla_{m}\nabla_{s}v^{l})\\
 & = & -h_{a}^{m}h_{b}^{n}h_{l}^{c}n^{s}\nabla_{m}n_{n}\nabla_{s}v^{l}+h_{a}^{m}h_{b}^{s}h_{i}^{c}v^{l}\nabla_{m}n^{i}\nabla_{s}n_{l}+h_{a}^{m}h_{b}^{s}h_{l}^{c}\nabla_{m}\nabla_{s}v^{l}\\
 & = & K_{ab}h_{l}^{c}n^{s}\nabla_{s}v^{l}+K_{a}^{c}K_{bl}v^{l}+h_{a}^{m}h_{b}^{s}h_{l}^{c}\nabla_{m}\nabla_{s}v^{l}
\end{array}.
\]
 Next, using the symmetry of the extrinsic curvature, one gets for
eq. (\ref{eq: ricci}) :
\begin{equation}
D_{a}D_{b}v^{c}-D_{b}D_{a}v^{c}=\left(K_{a}^{c}K_{bl}-K_{b}^{c}K_{al}\right)v^{l}+h_{a}^{m}h_{b}^{s}h_{l}^{c}\left(\nabla_{m}\nabla_{s}v^{l}-\nabla_{s}\nabla_{m}v^{l}\right)\label{eq:ricci2}
\end{equation}
Using $\nabla_{m}\nabla_{s}v^{l}-\nabla_{s}\nabla_{m}v^{l}={}^{(4)}R_{ims}^{l}v^{i}$
, eq. (\ref{eq: ricci}) :and eq. (\ref{eq:ricci2}) we find: 
\[
\left(K_{a}^{c}K_{bk}-K_{b}^{c}K_{ak}\right)v^{k}+h_{a}^{m}h_{b}^{s}h_{l}^{c}{}^{(4)}R_{kms}^{l}v^{k}=^{(3)}R_{kab}^{c}v^{k}
\]
and thus the Gauss relation for non-Cauchy foliation is 
\[
h_{a}^{m}h_{b}^{s}h_{l}^{c}h_{k}^{h}{}^{(4)}R_{hms}^{l}=^{(3)}R_{kab}^{c}-K_{a}^{c}K_{bk}+K_{b}^{c}K_{ak}.
\]
Note that this term is different from the Gauss relation for Cauchy
foliation. Using a non-Cauchy foliation leads to different signs for
the two last terms. 

If we contract the Gauss relation on the indices $c$ and $a$ and
use $h_{ma}h_{l}^{a}=h_{ml}=g_{ml}-n_{m}n_{l}$, we get: 
\begin{equation}
h_{a}^{s}h_{b}^{h}{}^{(4)}R_{hs}-n^{l}h_{a}^{s}h_{hb}{}^{(4)}R_{lsm}^{h}n^{m}=^{(3)}R_{ab}-KK_{ab}+K_{a}^{i}K_{ib}.\label{eq:contrc gaus eq}
\end{equation}

The next step is to derive $n^{l}h_{a}^{s}h_{hb}{}^{(4)}R_{lsm}^{h}n^{m}$
with is the Ricci identity applied to the vector $n^{a}$, and projecting
it twice onto $\Sigma_{\text{r}}$ and once along $n^{a}$:
\begin{equation}
h_{am}n^{i}h_{b}^{n}(\nabla_{n}\nabla_{i}n^{m}-\nabla_{i}\nabla_{n}n^{m})=h_{am}n^{i}h_{b}^{n}{}^{(4)}R_{jni}^{m}n^{j}\label{eq:einst begin}
\end{equation}
In order to calculate this term we work out $\nabla_{a}n^{b}$:
\[
K_{ab}=-\frac{1}{2}\mathcal{L}_{n}h_{ab}=
\]
\[
=-\frac{1}{2}\left(n^{i}\nabla_{i}h_{ab}+h_{ib}\nabla_{a}n^{i}+h_{ai}\nabla_{b}n^{i}\right)=
\]
\[
=-\frac{1}{2}\left(n^{i}\nabla_{i}\left(g_{ab}-n_{a}n_{b}\right)+\left(g_{ib}-n_{i}n_{b}\right)\nabla_{a}n^{i}+\left(g_{ai}-n_{a}n_{i}\right)\nabla_{b}n^{i}\right)=
\]

\[
=-\frac{1}{2}\left(\nabla_{a}n_{b}+\nabla_{b}n_{a}-n_{a}n^{i}\nabla_{i}n_{b}-n_{b}n^{i}\nabla_{i}n_{a}\right)
\]
Using $n_{a}=N\nabla_{a}r$ we find that $n^{i}\nabla_{i}n_{a}=-D_{a}lnN$
\footnote{$n^{i}\nabla_{i}n_{a}=n^{i}\nabla_{i}\left(N\nabla_{a}r\right)=n^{i}\nabla_{i}N\nabla_{a}r+Nn^{i}\nabla_{i}\nabla_{a}r=$$=N^{-1}n_{a}n^{i}\nabla_{i}N+Nn^{i}\nabla_{a}\nabla_{i}r=N^{-1}n_{a}n^{i}\nabla_{i}N+Nn^{i}\nabla_{a}(N^{-1}n_{i})=$$N-1(n_{a}n^{i}\nabla_{i}N-\nabla_{a}N)=-h_{ai}\nabla^{i}lnN=-D_{a}lnN$}we
find 
\[
K_{ab}==-\frac{1}{2}\left(\nabla_{a}n_{b}+\nabla_{b}n_{a}+n_{a}D_{b}lnN+n_{b}D_{a}lnN\right)
\]
Thus $\nabla_{[a}n_{b]}=-K_{ab}-n_{[a}D_{b]}lnN$

Returning to eq. (\ref{eq:einst begin})we get
\begin{eqnarray*}
\begin{array}{ccc}
h_{am}n^{i}h_{b}^{n}R_{jni}^{(4)m}n^{j}=h_{am}n^{i}h_{b}^{n}(\nabla_{n}\nabla_{i}n^{m}-\nabla_{i}\nabla_{n}n^{m})\\
=h_{am}n^{i}h_{b}^{n}\left[\nabla_{n}(-K_{i}^{m}-n_{i}D^{m}lnN)-\nabla_{i}(-K_{n}^{m}-n_{n}D^{m}lnN)\right]\\
=h_{am}n^{i}h_{b}^{n}[-\nabla_{n}K_{i}^{m}-\nabla_{n}n_{i}D^{m}lnN-n_{i}\nabla_{n}D^{m}lnN\\
+\nabla_{i}K_{n}^{m}-\nabla_{i}n_{n}D^{m}lnN-n_{n}\nabla_{i}D^{m}lnN]\\
=h_{am}h_{b}^{n}[K_{i}^{m}\nabla_{n}n^{i}-\nabla_{n}D^{m}lnN+n^{i}\nabla_{i}K_{n}^{m}-D_{n}lnN\cdot D^{m}lnN-n_{n}n^{i}\nabla_{i}D^{m}lnN]\\
=K_{ia}\nabla_{b}n^{i}-D_{b}D_{a}lnN+h_{am}h_{b}^{n}n^{i}\nabla_{i}K_{n}^{m}-D_{b}lnM\cdot D_{a}lnN\\
=-K_{ai}K_{b}^{i}-N^{-1}D_{b}D_{a}N+h_{am}h_{b}^{n}n^{i}\nabla_{i}K_{n}^{m}
\end{array}.
\end{eqnarray*}

Note we have used $K_{i}^{a}n^{i}=0$, $n^{i}\nabla_{a}n_{i}=0$,
$n_{i}n^{i}=1$, $n^{i}\nabla_{i}n_{a}=-D_{a}lnN$ and $h_{a}^{i}n_{i}=0$
to get the third equality. Let us now show that the term $h_{am}h_{b}^{n}n^{i}\nabla_{i}K_{n}^{m}$
is related to $\mathcal{L}_{r}K_{ab}$. Indeed 
\[
\mathcal{L}_{r}K_{ab}=r^{i}\nabla_{i}K_{ab}+K_{ib}\nabla_{a}r^{i}+K_{ai}\nabla_{b}r^{i}.
\]
Using $r_{a}=Nn_{a}+W_{a}$ and $\nabla_{[a}n_{b]}=-K_{ab}-n_{[a}D_{b]}lnN$
we get
\[
\mathcal{L}_{r}K_{ab}=Nn^{i}\nabla_{i}K_{ab}-2NK_{ib}K_{a}^{i}-K_{ib}n_{a}D^{i}N-K_{ai}n_{b}D^{i}N.
\]
Projecting on $\varSigma_{r}$ by applying $h^{mn}$on both side and
using $\mathcal{L}_{r}K_{ab}=h_{a}^{n}h_{b}^{m}\mathcal{L}_{r}K_{nm}$we
get
\[
\mathcal{L}_{r}K_{ab}=Nh_{a}^{n}h_{b}^{m}n^{i}\nabla_{i}K_{nm}-2NK_{ib}K_{a}^{i}.
\]
Thus 
\begin{equation}
h_{am}n^{i}h_{b}^{n}{}^{(4)}R_{jni}^{m}n^{j}=N^{-1}\mathcal{L}_{r}K_{ab}-N^{-1}D_{b}D_{a}N+K_{ai}K_{b}^{i}\label{eq:rimamOn hhnn}
\end{equation}
The left hand side of (\ref{eq:rimamOn hhnn}) is a term which appears
in the contracted Gauss equation (\ref{eq:contrc gaus eq}) . Therefore,
by combining the two equations, we get:
\begin{equation}
h_{a}^{s}h_{b}^{h}{}^{(4)}R_{hs}=N-1\mathcal{L}_{r}K_{ab}-N^{-1}D_{b}D_{a}N+{}^{(3)}R_{ab}-KK_{ab}+2K_{a}^{i}K_{ib}\label{eq:Ricci hh}
\end{equation}
Note that this term is different from the one we get using Cauchy
foliation. Using non-Cauchy foliation leads to different sign for
the first, and the last two last terms of the left hand side terms. 

Finally, contracting the $(3+1)D$ Einstein equation (\ref{eq:4d einst eq})
with the induced metric we get for the non-Cauchy surface foliation
: 
\begin{equation}
R_{ab}^{(3)}-KK_{ab}+2K_{ai}K_{b}^{i}+N^{-1}\left(\mathcal{L}_{r}K_{ab}-D_{a}D_{b}N\right)=8\pi\left(S_{ab}-\frac{1}{2}\left(S-P\right)h_{ab}\right),\label{eq:Estn3D-1}
\end{equation}
Where $D_{a}$ represent the 2+1 covariant derivatives, $S_{ab}=h_{ac}h_{ad}T^{cd}$
, $P=n_{c}n_{d}T^{cd}$ and $F_{a}=h_{ac}n_{b}T^{cb}$.

Next we find the constraint relevant for the non-Cauchy surface foliation. 

The first is obtained by projection of the Einstein equation along
$n^{a}$, i.e. the normal to the hypersurface $\varSigma_{r}$ . Contracting
eq. (\ref{eq:contrc gaus eq}) with $h^{ab}$ and using $h^{ab}{}^{(4)}R_{ab}=\left(g^{ab}-n^{a}n^{b}\right){}^{(4)}R_{ab}={}^{(4)}R-{}^{(4)}R_{ab}n^{a}n^{b}$
and $h^{ab}n^{l}h_{a}^{s}h_{hb}{}^{(4)}R_{lsm}^{h}n^{m}=n^{l}h_{h}^{s}{}^{(4)}R_{lsm}^{h}n^{m}=n^{l}{}^{(4)}R_{lm}n^{m}-n^{l}n_{h}n^{s}{}^{(4)}R_{lsm}^{h}n^{m}=n^{l}{}^{(4)}R_{lm}n^{m}$
we get 
\[
^{(4)}R-2{}^{(4)}R_{ab}n^{a}n^{b}=^{(3)}R-K^{2}+K^{ij}K_{ij}.
\]
Using $T=g_{ab}T^{ab}=h_{ab}T^{ab}+n_{a}n_{b}T^{ab}=S+P$ we find
$^{(4)}R=-8\pi(S+P)$ $^{(4)}R_{ab}n^{a}n^{b}=8\pi\left(P-\frac{1}{2}T\right)=4\pi\left(P-S\right)$
and thus we obtain the first constraint:
\[
^{(3)}R-K^{2}+K^{ij}K_{ij}=-16\pi P.
\]
Finally, in order to derive the second constraint we need to derive
the relevant Codazzi relation by applying the Ricci identity to $n^{a}$:
\[
\nabla_{m}\nabla_{s}n^{k}-\nabla_{s}\nabla_{m}n^{k}={}^{(4)}R_{ims}^{k}n^{i}.
\]
Projecting this onto $\Sigma_{r}$, we get 
\[
h_{a}^{m}h_{b}^{s}h_{k}^{c}{}^{(4)}R_{ims}^{k}n^{i}=h_{a}^{m}h_{b}^{s}h_{k}^{c}\left(\nabla_{m}\nabla_{s}n^{k}-\nabla_{s}\nabla_{m}n^{k}\right).
\]
Since $h_{b}^{s}h_{k}^{c}\nabla_{s}n^{k}=-K_{b}^{c}$, $h^{ab}=g^{ab}-n^{a}n^{b}$
and $h_{b}^{s}n^{m}n_{k}{}^{(4)}R_{ims}^{k}n^{i}=0$ we get after
contracting the indices $a$ and $c$ 
\[
h_{b}^{s}{}^{(4)}R_{is}n^{i}=-D_{i}K_{b}^{i}+D_{b}K.
\]
Finally, using Einstein equation once onto $\varSigma_{r}$ and once
along the normal $n^{a}$ we find: 

\subsubsection*{
\begin{equation}
D_{a}K-D_{i}K_{a}^{i}=8\pi F_{a}.\label{eq:constrnt2-1}
\end{equation}
}

\subsubsection*{To conclude:}

Where as foliating spacetime using time-like vector field $u^{a}$
leads to 

\begin{equation}
\widetilde{R}_{ab}^{(3)}+\widetilde{K}\widetilde{K}_{ab}-2\widetilde{K}_{ai}\widetilde{K}_{b}^{i}+N^{-1}\left(\mathcal{L}_{t}\widetilde{K}_{ab}+\widetilde{D}_{a}\widetilde{D}_{b}N\right)=8\pi\left(\widetilde{S}_{ab}-\frac{1}{2}\left(\widetilde{S}-E\right)\gamma_{ab}\right),\label{eq:Estn3D-1-1-1}
\end{equation}

\[
^{(3)}\widetilde{R}+\widetilde{K}^{2}-\widetilde{K}^{ij}\widetilde{K}_{ij}=16\pi E.
\]

\begin{equation}
\widetilde{D}_{i}\widetilde{K}_{a}^{i}-\widetilde{D}_{a}\widetilde{K}=8\pi p_{a}.\label{eq:constrnt2-1-1-1}
\end{equation}
where the lapse function $N$ and shift vector $U_{a}$ satisfy $t_{a}=Nu_{a}+U_{a}$
, $t^{a}\nabla_{a}t=1$ and $t$ is constant on $\widetilde{\Sigma}_{\text{t}}$and
the hyper-surface metric $\gamma_{ab}$ is given by $g_{ab}=\gamma_{ab}-u_{a}u_{b}$.
In this case the extrinsic curvature tensor of the hyper-surfaces
is given by $\widetilde{K}_{ab}=-\frac{1}{2}\mathcal{L}_{u}\gamma_{ab}$
and $\widetilde{D}_{a}$ represent the 3 spatial covariant derivatives,
$\widetilde{S}_{ab}=\gamma_{ac}\gamma_{ad}T^{cd}$ , $E=u_{c}u_{d}T^{cd}$
and $\text{p}_{a}=-\gamma_{ac}u_{b}T^{cb}$.

Foliating along a spacelike vector fields leads to somewhat different
equations:
\begin{equation}
R_{ab}^{(3)}-KK_{ab}+2K_{ai}K_{b}^{i}+N^{-1}\left(\mathcal{L}_{r}K_{ab}-D_{a}D_{b}N\right)=8\pi\left(S_{ab}-\frac{1}{2}\left(S-P\right)h_{ab}\right),\label{eq:Estn3D-1-1}
\end{equation}

\[
^{(3)}R-K^{2}+K^{ij}K_{ij}=-16\pi P.
\]

\begin{equation}
D_{a}K-D_{i}K_{a}^{i}=8\pi f_{a}.\label{eq:constrnt2-1-1}
\end{equation}
where $D_{a}$ represent the 2+1 covariant derivatives, $S_{ab}=h_{ac}h_{ad}T^{cd}$
, $P=n_{c}n_{d}T^{cd}$ and $f_{a}=-h_{ac}n_{b}T^{cb}$.

\end{document}